\documentclass{iau}
\usepackage{graphicx}

\newcommand{\apj}{ApJ}           % Astrophysical Journal
\newcommand{\apjl}{ApJ}           % Astrophysical Journal
\newcommand{\aap}{A\&A}
\newcommand{\araa}{ARA\&A}

\newcommand{\pasp}{PASP}
\def\msun{$M_\odot$}

\title[Local Star Formation] %% give here short title %%
{Star Formation in the Local Milky Way}

\author[]   %% give here short author list %%
{Charles J. Lada$^1$}

\affiliation{$^1$Smithsonian Astrophysical Observatory\\email: {\tt clada@cfa.harvard.edu}\\}
\pubyear{2015}
\volume{314}  %% insert here IAU Symposium No.
\pagerange{119--126}
\jname{Young Stars \& Planets Near the Sun}
\editors{J. H. Kastner, B. Stelzer, \& S. A. Metchev, eds.}
\begin{document}

\maketitle

\begin{abstract}
Studies of molecular clouds and young stars near the sun have provided invaluable insights into the process of star formation. Indeed, much of our physical understanding of this topic has been derived from such studies. Perhaps the two most fundamental problems confronting star formation research today are: 1) determining the origin of stellar mass and 2) deciphering the nature of the physical processes that control the star formation rate in molecular gas.  As I will briefly outline here, observations and studies of local star forming regions are making particularly significant contributions toward the solution of both these important problems. 
%% add here a maximum of 10 keywords, to be taken form the file <Keywords.txt>
\end{abstract}

\firstsection % if your document starts with a section,
              % remove some space above using this command.
\section{Introduction}

In this contribution I will describe what I consider to be two of the most fundamental problems that confront star formation research and how observations of the local Milky Way are providing critical insights into their resolution. These two basic questions are: 1) what is the origin of stellar mass? and 2) what are the physical processes that control the rate at which the gaseous interstellar medium (ISM) transforms a significant part of itself into stellar form? 

Stars are the fundamental objects of the astronomical universe.
They convert hydrogen, the primary product of the big bang, into the heavy elements of the periodic table. 
Through stellar evolution they control the evolution of all stellar systems from  stellar clusters to the largest galaxies.
They provide the sites for planetary systems and the energy necessary for the development and existence of life.
Consequently, knowledge of their origins is both of compelling intrinsic interest as well as essential for understanding the closely related problems of planet formation and galaxy formation and evolution.

\section{A Predictive Theory of Star Formation}

In my view the ultimate goal of star formation research is the development of a predictive theory of stellar origins. By that, I mean the construction of a theory that, given a limited set of initial conditions, is able to predict the basic physical properties of stars, the rate at which these stars form, and how this rate varies in time.  Probably the most fundamental physical properties of stars that should be predicted by any respectable theory of stellar origins are their compositions, luminosities, temperatures, size,s and masses. The first of these properties is trivially predicted since we have {\it a priori} empirical knowledge of the composition of the ISM from which the stars themselves form. Additionally, the beautiful and powerful theory of stellar structure and evolution, developed in the twentieth century, can already predict the luminosities, temperatures, and sizes of stars, once their initial masses and compositions are specified.  This leaves stellar mass as the one fundamental stellar property for which we have yet no physical explanation and which a star formation theory first and foremost must explain. Star formation rates are known to vary considerably within and between galaxies (e.g., \cite[Kennicutt 1998]{1998ARAA}) and between local molecular clouds (e.g., \cite[Lada et al. 2010]{2010ApJ...724..687L}). However, little is understood about the physical processes that control the star formation rate (SFR) in interstellar gas. These processes and how they vary in space and time are what control the evolution of all stellar systems from star clusters to galaxies. In doing so they drive cosmic evolution itself. A general theory of star formation must be able to predict the SFR and its environmental and temporal variations in molecular clouds and galaxies. 

\section{The Origin of Stellar Mass}

\subsection{The Stellar Initial Mass Function }

According to the theory of stellar structure and evolution, once formed the subsequent life history of a star is entirely predetermined by the only two parameters: the star's initial mass and, to a lesser extent, its chemical composition.
The frequency distribution of stellar masses at birth, or the initial mass function (IMF) of stars, thus plays a pivotal role in the evolution of all stellar systems. In the absence of a general theory that can predict stellar masses and the IMF, these must be first determined empirically. The first attempt to determine the IMF was by \cite[Salpeter (1955)]{es55} more than half a century ago. He derived the IMF from the luminosity function of local field stars (i.e., stars within $\sim$ 500 pc of the sun) by converting stellar luminosities to masses using  empirical mass-luminosity relations. He corrected for losses in the stellar number counts due to stellar evolution by assuming a constant SFR over the age of the Galaxy. Salpeter demonstrated that between 1 and 10 \msun \ the IMF had the form of a power-law with an index of -1.3 (when adopting the conventional logarithmic mass binning for this function). 

Numerous subsequent determinations of the local field star IMF increased the range of distances and masses over which the IMF could be determined, with masses now ranging from OB stars to the hydrogen burning limit and somewhat below (e.g., \cite[Scalo 1978]{js78}, \cite[Kroupa 2002]{2002Sci...295...82K}, 
\cite[Chabrier 2003]{2003PASP..115..763C}). In addition, infrared studies of local, very young star clusters provided independent determinations of the IMF over an even larger dynamic range in mass, from the deuterium burning limit to OB stars (e.g., \cite[Muench et al. 2002]{2002ApJ...573..366M}). These studies have found 
that the shape of the IMF is lognormal-like and exhibits a broad peak between 0.1 and 0.5 \msun\, suggesting a characteristic mass associated with star formation of about 0.25 \msun. Less than $\sim$ 25 \% of the objects formed in the Trapezium star forming event were found to have substellar masses (\cite[Muench et al. 2002]{2002ApJ...573..366M}).  Moreover, the fact that the form of the field star IMF so closely matches that of an embedded cluster such as the Trapezium in Orion (see Figure \ref{fig:cmfimf}) is quite profound, since field stars were formed over billions of years of Galactic history and over a large volume of space (perhaps as much as a kpc in extent) while an embedded cluster like the Trapezium formed its stars in only a few million years in a volume perhaps ten million times smaller. This  indicates that the functional form of the IMF in the disk of the Milky Way is very likely universal in both space and time.

Another stellar property that should be met by a predictive theory of star formation is that of stellar multiplicity. It is well established that many stars in the Milky Way are contained in multiple systems which are mostly binaries. In recent years it has been shown that stellar multiplicity is a strong function of spectral type and stellar mass. The {\it single} star fraction ranges from only about 1-20\%  for OB stars to as much as 70-80 \% for M stars and objects near the hydrogen burning limit. Since about 75\% of the stars that make up the standard IMF are M stars, it is apparent that {\it the typical outcome of the star formation process is a single M dwarf star} (\cite[Lada 2006]{2006ApJ...640L..63L}).

\subsection{The Dense Core Mass Function}

\begin{figure}[] 
 \vspace*{-2.0 cm}
\begin{center}
\includegraphics[width=0.7\columnwidth]{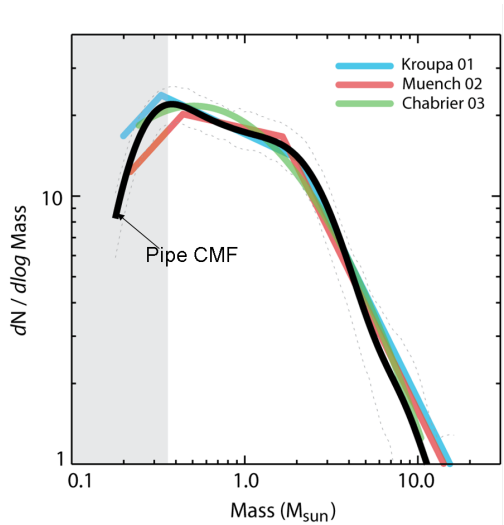} 
 \vspace*{-2.5 cm}
 \caption{Comparison between the CMF of the Pipe molecular cloud and the IMFs derived for field stars and the young Trapezium cluster in Orion. The IMFs have all been scaled (i.e., shifted) higher in mass by a factor of 3 to match the Pipe CMF. The close similarity in shape between the CMF and IMF upon shifting the latter function suggests that the IMF derives directly from the CMF with a star formation efficiency of approximately 30\%.}
 \label{fig:cmfimf}
\end{center}
\end{figure}

Observations of the cold ISM at infrared and millimeter wavelengths indicate that stars form from dense molecular filaments, clumps and cores in cold molecular clouds (e.g., \cite[Lada 1992]{1992ApJ...393L..25L}, \cite[Beichman et al. 1986]{1986ApJ...307..337B}, \cite[Yun \& Clemens 1990]{1990ApJ...365L..73Y}, \cite[Lada et al. 2010]{2010ApJ...724..687L}). The physical conditions in this dense molecular gas can be thought of as the initial conditions for a predictive theory of star formation. These conditions would include the masses, sizes, temperatures, densities, pressures, kinematics, and compositions of dense  cores, particularly starless ones. Studies of nearby ( $<$ 150 pc) local clouds have provided a wealth of information concerning these cores and the process of star formation within them. For the purposes of this paper I will now only describe what we know of dense core masses, their mass spectrum and its relation to the stellar IMF. Investigations of the (dense) core mass function, or CMF, have a much more recent history than studies of the IMF. Only in the last decade or so have improved observational capabilities (e.g., wide-field, near-infrared extinction mapping, millimeter-wave detector arrays, space-based far-infrared and sub-millimeter observations, increased computational power, etc.) enabled measurements of dense core masses over a sufficiently large dynamic range to make useful comparisons to the IMF. Earlier observations of nearby clouds suggested that the CMF could be described as a power-law with general similarity to the IMF  and with a hint at a flattening toward lower masses (e.g., $\rho$ Ophiuchi: \cite[Motte et  al. 1998]{1998A&A...336..150M}). Later, more sensitive observations were able to measure a clear departure from a power-law form at low core masses as well the presence of a broad peak in the CMF (e.g., Pipe Nebula: \cite[Alves et al. 2007]{2007A&A...462L..17A}; Aquila Rift: \cite[Andr{\'e} et al. 2010]{2010A&A...518L.102A}). These observations showed that the shapes of the CMF and IMF were quite similar, as can be observed in Figure \ref{fig:cmfimf}. However the CMF was found to peak at higher masses ($\sim$ 1--2 \msun) than the IMF. {\it These facts suggest that the IMF directly derives from the CMF with an efficiency of $\sim$ 25-30\%} (\cite[Alves et al. 2007]{2007A&A...462L..17A}). Individually, dense cores appear to be the direct progenitors of stars. 

If the above considerations are correct, then the question of the origin of stellar masses and the IMF becomes the question of the origin of the CMF and dense core masses.
The origin of core masses is tied to the process of molecular cloud fragmentation and thus to the evolution of cloud structure. The characteristic mass of the CMF likely represents a characteristic mass scale for cloud fragmentation.
In the nearby Pipe Nebula, \cite[Lada et al. (2008)]{2008ApJ...672..410L} noted that the characteristic mass of the CMF was approximately the critical Bonnor-Ebert mass for core population in that cloud. This suggests that, at least in the Pipe Nebula, the CMF originated from a Jeans-like process of thermal fragmentation in a pressurized medium. Whether or not such an explanation can account for the core formation process in other clouds and environments remains to be seen. However, it is interesting to consider that if cores at the peak of the CMF have masses comparable to the critically stable, Bonnor-Ebert mass, they will predominately form single stars once they become unstable and collapse. Moreover, with typical star formation efficiencies of 20-30\%,  such cores will produce stars whose final masses will have values near the observed peak of the IMF. 

\begin{figure}[] 
% \vspace*{-2.0 cm}
\begin{center}
\includegraphics[width=0.75\columnwidth,angle=-90]{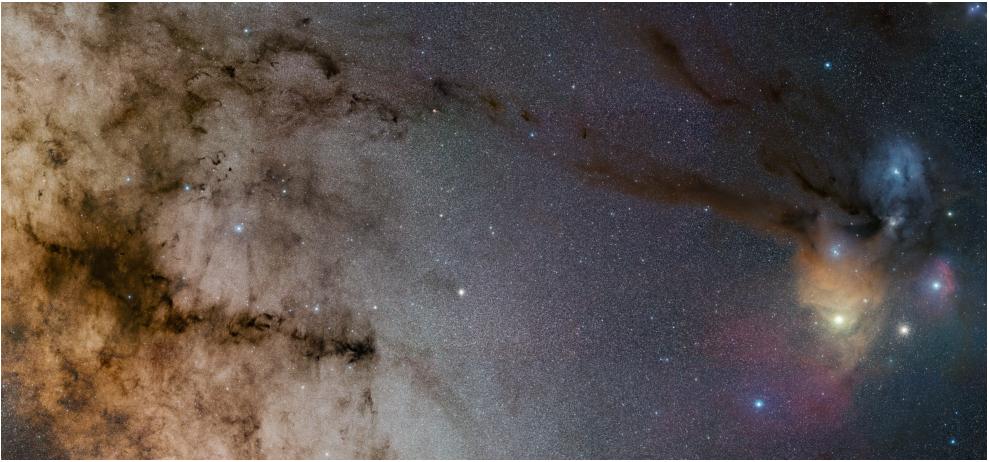} 
 \vspace*{-1.5 cm}
 \caption{The Pipe (left) and Ophiuchus (right) dark nebulae. These nearest examples of massive, cold molecular clouds are at the same distance and are roughly similar in mass and extent, yet they display dramatically different levels of star formation activity. {\bf Credit}: ESO/S. Guisard}
 \label{fig:pipeoph}
\end{center}
\end{figure}

\section{The Star Formation Rate}

Figure \ref{fig:pipeoph} is a stunning image and is one of my favorites. It contains two of the most intriguing local  molecular clouds, the Pipe Nebula and the Ophiuchi cloud.  Located at nearly the same distance (D $\approx$ 125 pc), in nearly the same volume of Galactic space, these clouds are of very similar mass ($\sim$ 10$^{4}$ \msun) and size (L $\sim$ 15-20 pc). Yet, as the image clearly shows, they differ substantially in their level  of star formation activity. The Ophiuchi cloud is ablaze in star formation while the Pipe appears cold, dark and quiescent. How is it that two clouds so similar in mass, size, and age and in such close proximity, can be so different in their star formation? 
Interestingly, the ages of the stellar populations in the two clouds are essentially the same (2-3 Myr; \cite[Covey et al. 2010]{,2010ApJ...722..971C}).  Given the universal nature of the functional form of the IMF,  the difference in stellar content and appearance of the two clouds must therefore be  due to drastically differing rates of star formation within them. 

Traditionally, researchers who study star formation in the Milky Way have not devoted much attention to the question of understanding the SFR in molecular gas. But for extragalactic astronomy, the SFR plays a critical role. For example, it is the primary tracer or metric used to describe the evolution of galaxies over cosmic time (e.g, \cite[Madau et al. 1998]{1998ApJ...498..106M}) and to investigate the connection between the level of star formation and the gas content of galaxies (\cite[Kennicutt 1998]{1998ARAA}). It has become increasingly apparent to me that the SFR also plays a critical role in GMCs and knowledge of the factors that control the SFR in local molecular clouds is necessary for the development of a comprehensive and predictive theory of star formation. Moreover, understanding the underlying physics that controls the SFR in local molecular gas is likely a key stepping stone to understanding  the nature and evolution of galaxies as well. 

To address this issue we will take another look at the stellar IMF. Over the past 50 years research on the IMF has almost exclusively dealt with measuring its functional form, its extent in mass, and the degree to which these quantities vary in space and time. However, it is also instructive to consider what information can be gleaned from the amplitude or normalization of the IMF.  As an illustrative example, let us consider the log-normal form of the IMF for a single stellar population, like a young cluster. We can write the IMF as
\begin{equation}
\xi(log m_*) = C_0 {\rm exp}[-({\rm log}(m_*/m_c))^2/2\sigma^2].
\end{equation}

\noindent
Here $m_*$ is the stellar mass, $m_c$ is the characteristic mass, and $\sigma$ is the width of the log normal function. The normalization coefficient is given by $C_0 = N_*(2 \pi)^{-1/2}\sigma^{-1}$, where $N_*$ is the total number of stars formed in the cluster (or stellar population) being considered. This coefficient and thus the IMF are directly connected to the SFR since $N_* = b_* \Delta \tau_{sf}$ and SFR $= b_* <m_*>$, where $b_*$ is the average birthrate (yr$^{-1}$) of stars in the cluster, $\Delta\tau_{sf}$ the duration of star formation in the cluster (yr) and $<m_*>$ the average mass (\msun ) of a star in the cluster. 
The central point here, however, is that in local regions of star formation $N_*$ and $\Delta\tau_{sf}$ are observable quantities and {\it we can directly measure the SFR in local clouds}. This is fundamentally different than the situation in more distant regions and in galaxies where the SFR has to be determined indirectly, usually requiring  the aid of population synthesis models (e.g., \cite[Kennicutt  1998]{1998ARAA}).

We can now quantify the absolute and relative SFRs in the clouds of Figure \ref{fig:pipeoph}. There are over 300 young stellar objects (YSOs) within the Ophiuchi cloud (e.g., \cite[Wilking et al. 2008]{2008hsf2.book..351W}), while only 21 are found in the Pipe (\cite[Forbrich et al. 2009]{2009ApJ...704..292F}; \cite[Forbrich et al. 2010]{2010ApJ...719..691F}). Since the ages of the stellar populations in the two clouds are essentially the same their star formation rates must differ by a factor of $\approx$ 15. These two objects are not the only clouds to display such large variations in SFRs. Based on complete and accurate measurements of the masses and stellar contents of a nearly complete sample of molecular clouds within 0.5 kpc of the sun, \cite[Lada et al. (2010)]{2010ApJ...724..687L} demonstrated that the specific star formation rates (i.e., the SFR per unit cloud mass or sSFR) of local clouds vary by more than an order of magnitude. This variation is independent of cloud mass over a range of two orders of magnitude in cloud mass, a result hinted at in early CO observations of more distant and massive Milky Way clouds (\cite[Mooney \& Solomon 1988]{1988ApJ...334L..51M}).

\begin{figure}[]
\begin{center}
\vspace*{-1.5 cm}
\includegraphics[width=0.6\columnwidth]{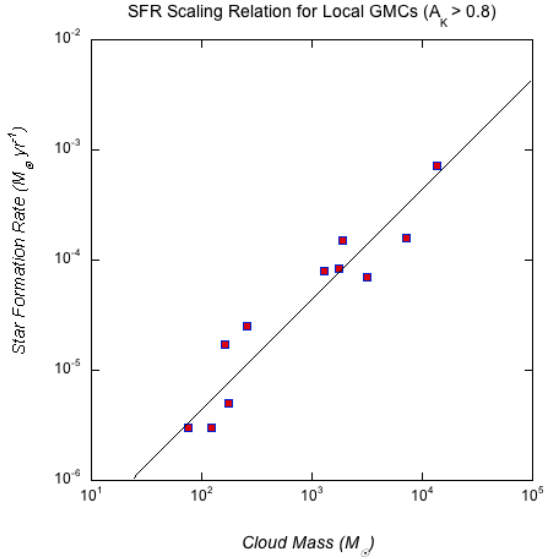} 
\vspace*{-1.5cm}
\caption{The scaling relation between SFR and high extinction (i.e., A$_K \geq$ 0.8 magnitudes), presumably dense, gas mass for local GMCs.  Adapted from Lada et al. (2010). See text.}
\label{fig:sfrvsdg}
\end{center}
\end{figure}

That the SFRs of local GMCs can be so accurately measured and are found to vary so significantly is an exciting development for star formation research.  This is because we can measure the basic physical properties of the local clouds in exquisite detail. Thus astronomers are in an excellent position to determine those physical properties that set the SFR and decipher the processes that control the SFR in molecular gas. 

Earlier observations of the Orion B GMC have long suggested that star formation occurs almost exclusively in extended regions of dense gas (e.g., Lada 1992). In their study of the SFRs in the local cloud sample, \cite[Lada et al. (2010)]{2010ApJ...724..687L} discovered a relatively tight scaling relation between the mass of high opacity gas and the global SFR. This scaling relation is shown in Figure \ref{fig:sfrvsdg}.
 This scaling is a linear power-law relation, that is,  SFR~$\propto$ M$_{dense}$. This has been interpreted to indicate that the SFR in a cloud is directly controlled by the amount of high extinction (and presumably dense) material contained within it. Indeed, recent observational studies of the Orion A GMC (\cite[Lombardi et al. 2014]{2014A&A...566A..45L}) and the {\it Spitzer} C2D local sample of dark clouds (\cite[Evans et al. 2014)]{2014ApJ...782..114E}; \cite[Heiderman \& Evans 2015]{2015ApJ...806..231H}) have shown that roughly 90\% of the protostellar objects in these clouds are found projected on high opacity (i.e., A$_K$ $\geq$ 0.8 magnitudes) material, providing further evidence of the tight relation between extended dense material and star formation.
 
 The linear scaling between dense gas mass and the SFR in local clouds is reminiscent of a similar global relation found for galaxies by Gao and Solomon (2004) who showed that FIR luminosities of galaxies (including disk galaxies and nuclear starbursts) were linearly correlated with the luminosities of HCN molecular-line emission. The FIR luminosity is a proxy for the SFR  and HCN emission is a tracer of the dense (n$_{H_2}$ $>$ 10$^4$ cm$^{-3}$) component of molecular gas. Subsequent observations by \cite[Wu et al. (2005)]{Wetal05}  comparing FIR and HCN luminosities of massive GMCs in the Milky Way also showed a linear correlation between the two quantities and this relation was found to extrapolate smoothly to that found by \cite[Gao \& Solomon (2004)]{gs04}  for galaxies, spanning a range in scale of over nine orders of magnitude. Moreover, \cite[Lada et al. (2012)]{2012ApJ...745..190L} found that the local GMCs also fit on this relation (after application of the appropriate calibrations for the conversion of FIR and HCN luminosities to SFRs and gas masses, respectively).
 
 The scaling relationship between the SFR and dense molecular gas mass found in local GMCs also appears to characterize star formation in galaxies. This similarity suggests that we are observing a similar physical process in star forming environments across all spatial scales. The linear scaling between SFR and dense gas mass suggests that the rate of star formation is directly controlled by the amount of dense gas that can be assembled in any star forming region.
Thus, the physical process that controls the assembly of dense material in a cloud and dictates the dense gas fraction likely also controls the cloud's SFR. As mentioned earlier, the evolution of the internal structure of this dense material is also likely responsible for generating the CMF and thus ultimately the stellar IMF.

\section{Concluding Remarks}

In the previous paragraphs of this contribution I have presented a  synopsis of two of the most fundamental problems confronting modern star formation research and have reviewed how studies of the local region of the Milky Way have produced critically important information that has led to some progress toward understanding these issues. I would be remiss, however, not to include some discussion of gains achieved in our understanding of the star formation process that pertain more directly to the subjects of this conference, young stars and in particular small moving groups of young stars near the sun. I will conclude this contribution with the following brief discussion conerning the origin of these small, local stellar moving groups.

\begin{figure}[]
\begin{center}
%\vspace*{-1.0 cm}
\includegraphics[width=0.5\columnwidth]{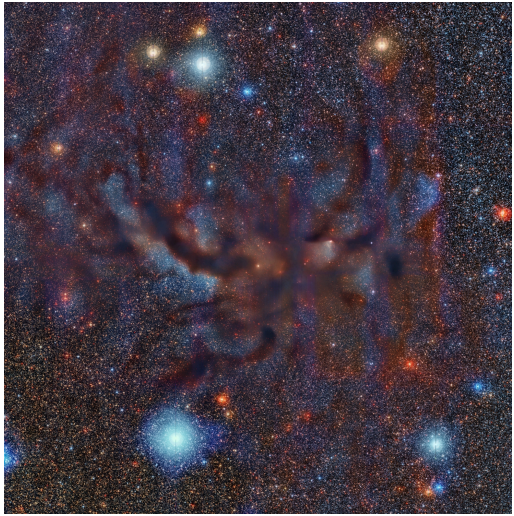} 
\vspace*{-0.5cm}
\caption{Optical Image of B 59, the most massive dense core in the Pipe Nebula. With 12 YSOs buried in this core, it is the most active site of star formation in the Pipe Nebula and an excellent candidate for producing a future moving group of young stars in the vicinity of the sun. (Image courtesy of J. Alves)}
\label{fig:b59}
\end{center}
\end{figure}

Stars form in dense gas with a star formation efficiency (SFE $=  {M_*\over M_* + M_{gas}}$) of about 30\% in dense cores, and about 10\% in the more extended dense gas containing dense cores. Since only about 1-10\% of the mass of GMCs is in the form of dense gas, the global SFEs of GMCs can range between 0.1 - 2\% but are typically observed to be on the order of 1-2 \%. For the local sample of \cite[Lada et al. (2010)]{2010ApJ...724..687L}, $<$SFE$>$ $=$ 1\% $\pm$ 0.8\%. Stars and stellar groups form in bound regions of GMCs where the majority of the binding mass is gaseous. With such low efficiencies, cloud dispersal results in the production of unbound stellar groups (\cite[Lada 1987]{1987IAUS..115....1L}). This can be seen from simple application of the virial theorem, that is, ${\rm M_{tot} v^{2} = GM_{tot}^{2}/R}$, where ${\rm M_{tot} = M_* + M_{gas}}$. Initially, the stars have the virial velocities of the system, that is,  ${\rm v_* =  v = (GM_{tot}/R)^{1/2}}$, and if the gas is quickly removed from the system the escape velocity for the stars is ${\rm v_{esc} =  (GM_*/R)^{1/2}}$. So the system cannot remain bound unless $\rm v_* < v_{esc}$ which in turn requires SFE $>$ 50\%. This is in essence why GMCs generally spawn unbound expanding OB associations. The stellar expansion velocities resulting from this process are $\approx {\rm v_*}$. Bound or loosely bound stellar clusters can emerge from the more massive dense cloud cores containing embedded clusters or even small stellar groups where the SFE can reach 20-30\% within the volume in which the stellar groups have formed and where the gas removal is more adiabatic (e.g., \cite[Lada et al. 1984]{1984ApJ...285..141L}). 

Do we know of any dense cores in nearby clouds that are producing small clusters that could be the progenitors of local moving groups of young stars? I would like to propose here one possible candidate source. It is  known as Barnard 59 and is the most massive core in the Pipe Nebula. An optical image of this core is shown in Figure \ref{fig:b59}. Its mass is about 20 \msun \ (\cite[Rom{\'a}n-Z{\'u}{\~n}iga et al. 2009)]{2009ApJ...704..18} and it contains a small cluster of 12 YSOs (\cite[Brooke et al. 2007]{2007ApJ...655..364B}) with a total stellar mass of approximately 6-8 \msun. The core SFE $=$ 23-28\%.   Since there are no massive stars forming in this core, its disruption (possibly by the generation of outflows from the embedded stars) is likely not to be very violent. These conditions are ideal for the production of a small, loosely bound group of stars when the cluster emerges following the anticipated gradual and adiabatic-like dissipation of the cloud core.

\end{document}